\documentclass[
reprint, 
superscriptaddress,
 amsmath,amssymb,
 aps,  
prb,
floatfix,
]{revtex4-2}

\setcitestyle{super}

\usepackage[utf8]{inputenc}
\usepackage[T1]{fontenc}
\usepackage{graphicx}
\usepackage{dcolumn}
\usepackage{bm}
\usepackage{braket}
\usepackage[dvipsnames]{xcolor}

\renewcommand{\deg}{^{\circ}}

\begin{document}

\preprint{APS/123-QED}

\title{\textbf{Competition between contrast and laser noise \\ in laser threshold magnetometry using an external cavity diode laser} 
}

\author{Shao Qi Lim}
\email{Corresponding author: qi.lim@rmit.edu.au}
\affiliation{Department of Physics, School of Science, RMIT University, Melbourne 3001 VIC Australia.}

\author{Alexander A. Wood}
\affiliation{School of Physics, The University of Melbourne, Parkville 3010 VIC Australia.}

\author{Brett C. Johnson}
\email{Contact author: brett.johnson2@rmit.edu.au}
\affiliation{Department of Physics, School of Science, RMIT University, Melbourne 3001 VIC Australia.}

\author{Qiang Sun}
\affiliation{Department of Physics, School of Science, RMIT University, Melbourne 3001 VIC Australia.}

\author{Jan Jeske}
\affiliation{Fraunhofer Institute for Applied Solid State Physics IAF, Tullastraße 72, Freiburg, im Breisgau 79108 Germany.}

\author{Hiroshi Abe}
\affiliation{National Institutes for Quantum Science and Technology: QST, Takasaki, Gunma, 370-1292, Japan.}

\author{Takeshi Ohshima}
\affiliation{National Institutes for Quantum Science and Technology: QST, Takasaki, Gunma, 370-1292, Japan.}
\affiliation{Department of Materials Science, Tohoku University, Aoba, Sendai, Miyagi, 980-8579, Japan.}

\author{David J. Ottaway}
\affiliation{School of Physics, Chemistry and Earth Sciences, and Institute for Photonics and Advanced Sensing, Adelaide University, Adelaide SA 5005 Australia.}

\author{Heike Ebendorff-Heidepriem}
\affiliation{School of Physics, Chemistry and Earth Sciences, and Institute for Photonics and Advanced Sensing, Adelaide University, Adelaide SA 5005 Australia.}

\author{Robert E. Scholten}
\affiliation{School of Physics, The University of Melbourne, Parkville 3010 VIC Australia.}
\affiliation{MOGLabs, Carlton 3053 Australia.}

\author{Andrew D. Greentree}
\affiliation{Department of Physics, School of Science, RMIT University, Melbourne 3001 VIC Australia.}

\author{Brant C. Gibson}
\affiliation{Department of Physics, School of Science, RMIT University, Melbourne 3001 VIC Australia.}


\begin{abstract}
The negatively charged nitrogen-vacancy center (NV$^-$) in diamond is a versatile platform for quantum magnetometry under ambient conditions. Recently, laser threshold magnetometry (LTM) has been proposed as a means to significantly enhance the sensitivity of NV-based magnetometers by incorporating a diamond hosting NV$^-$ centers within a laser cavity and operating near threshold.   
While demonstrations have validated the concept, practical implementations remain technically demanding, requiring high pump powers and precise alignment of free-space cavities. It remains unclear whether the benefits of operating near threshold will outpace increased laser noise.
In this work, we integrate an NV-diamond with a high NV$^-$ content into a compact external cavity diode laser and demonstrate singlet infrared absorption optically detected magnetic resonance (ODMR). We report a five-fold enhancement in the ODMR contrast by operating near the laser threshold. However, the magnetic-field sensitivity is a factor of two worse near threshold ($17~\mathrm{nT/\sqrt{Hz}}$) than when operating far above threshold ($7.6~\mathrm{nT/\sqrt{Hz}}$). Consequently, on this platform, near-threshold operation currently yields a net negative effect on sensitivity because the threshold-enhanced laser noise outweighs the threshold-enhanced contrast. These results demonstrate that managing this contrast-noise trade-off is a critical challenge that must be addressed to realize the potential benefits of laser threshold magnetometry.
\end{abstract}


\maketitle

\section{\label{sec:intro}Introduction}

The negatively charged nitrogen-vacancy center (NV$^-$, hereafter NV) in diamond is one of the most extensively investigated solid-state spin systems with applications in quantum sensing and quantum information processing. \cite{taylor2008high, pezzagna2021quantum} Magnetometers based on NV enable nanoscale magnetic-field sensing in material science, nanotechnology, neurobiology and medicine. \cite{doherty2013nitrogen, schirhagl2014nitrogen, rondin2014magnetometry} 

NV-based magnetometers typically rely on photoluminescence (PL)-based optically detected magnetic resonance (ODMR), where magnetic-field variations are encoded in changes to the red fluorescence emission ($637-800$~nm). The achievable sensitivity is fundamentally constrained by the $\sim30$\% spin-dependent inter-system crossing (ISC) branching ratio, spin dephasing from environmental defects such as nitrogen, unwanted occupation of the NV$^0$ charge state, and limited photon collection efficiency. \cite{barry2020sensitivity} Absorption ODMR either via the NV singlet transition at 1042~nm \cite{acosta2010broadband, jensen2014cavity} or the triplet pump excitation ($520-532$~nm) \cite{ahmadi2017pump, ahmadi2018nitrogen} provides an alternative readout modality with improved photon collection efficiency due to the collimated nature of the transmitted probe beam. Laser threshold magnetometry (LTM) has been proposed as a means to further enhance ODMR contrast with high photon collection efficiency. Incorporating NV-diamonds within laser cavities introduces magnetic-field-dependent optical gain or loss, creating relatively large variations in output power when a laser is operated near threshold. \cite{jeske2016laser, dumeige2019infrared, webb2021laser}

Singlet infrared absorption-based ODMR exploits population dynamics within the NV singlet manifold, enabling magnetic resonance detection through changes in the transmission of a 1042~nm probe laser. The small absorption cross section ($\sigma \sim 2\times10^{-22}$ cm$^2$) results in low intrinsic contrast for typical NV ensembles measured under single-pass configurations ($<1\%$ at room temperature \cite{dumeige2013magnetometry} and a few percent at cryogenic temperatures\cite{acosta2010broadband}). This limitation has motivated the use of optical cavities and extended path lengths to enhance the effective interaction volume. \cite{jensen2014cavity, chatzidrosos2017miniature, schall2025high, tayefeh2025towards} LTM provides an alternative route by leveraging the nonlinear response of a laser near threshold, which can significantly amplify the ODMR contrast to near unity. \cite{jeske2016laser, dumeige2019infrared} Sensitivities in the $\mathrm{fT/\sqrt{Hz}}$ range pertinent to biological sensing and magneto-encephalography applications are projected. \cite{brookes2022magnetoencephalography, broser2018optically, elzenheimer2020magnetic}

While recent singlet absorption-based LTM experiments have demonstrated promising shot-noise limited sensitivities in the $\mathrm{fT/\sqrt{Hz}}$ range, \cite{schall2025laser} they remain constrained by high system complexity. Implementations of LTM to date have largely relied on optically pumped vertical external cavity surface-emitting lasers (VECSELs), which require high optical pump powers (several watts) and free-space cavities that are comparatively large and more sensitive to alignment drifts. \cite{lindner2024dual, gottesman2024infrared, rottstaedt2025two} These factors limit the practical implementation of LTM systems. While contrast enhancements from operating near threshold result in improved projected shot-noise limits, it remains unclear how increased laser noise near the threshold will influence the achievable sensitivity.

In this work, we demonstrate singlet infrared absorption ODMR in a compact external cavity diode laser (ECDL) incorporating an NV-diamond. The primary aim of this work is to evaluate platform viability and to systematically investigate the threshold dynamics, fundamental limits, and trade-offs of LTM systems. Compared to optically pumped VECSEL-based approaches, our ECDL is a compact and mechanically robust cavity with superior alignment stability. The laser gain medium is electrically pumped and requires only $40-130$~mW of electrical power under typical operating conditions ($25-80$~mA, $1.5-1.6$~V). We examine how the NV-diamond modifies the laser threshold and how the ODMR contrast and magnetic-field noise change near the laser threshold. We interpret the observations using established phenomenological models. These results establish a compact and mechanically robust alternative for NV-based LTM and identify the key physical mechanisms that currently limit performance in practical LTM systems.

\section{NV ODMR and LTM}

\begin{figure}[t]
\includegraphics[width=1.0\linewidth]{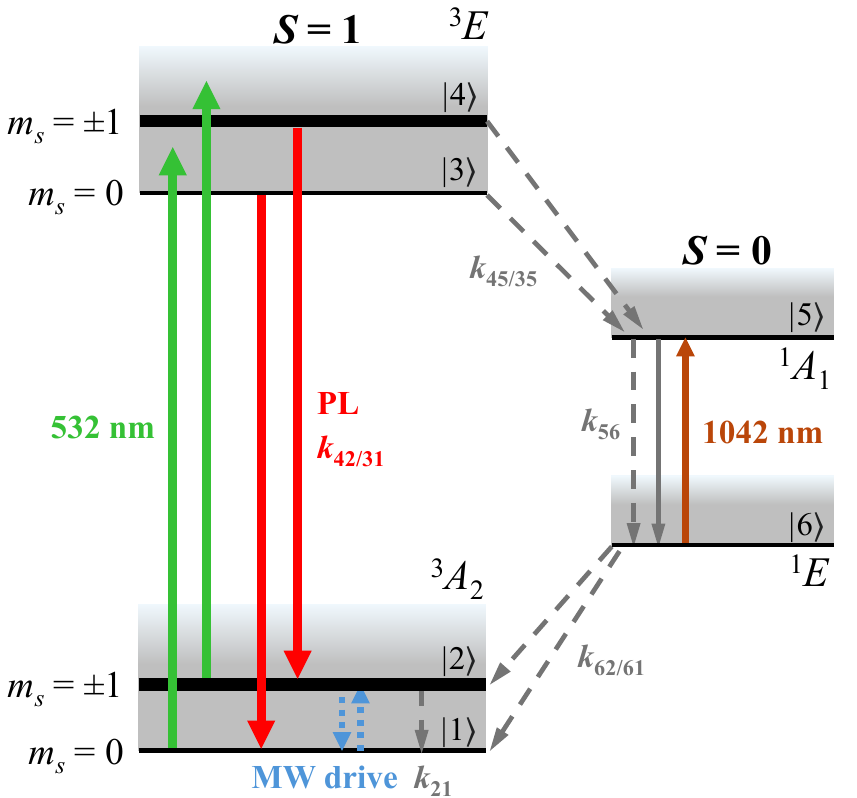}
\caption{\label{fig:NV-levels} Diamond NV energy levels at zero magnetic-field. The gray bands above each state represent a continuum of vibronically excited states responsible for off-resonant green absorption and PSB emission. }
\end{figure}

At zero magnetic-field, the energy level structure of the NV electronic spin defect ($S=1$) in diamond is shown in Fig.~\ref{fig:NV-levels}. \cite{doherty2012theory, rogers2015singlet} In the triplet ground state manifold ($^3A_2$), the $m_s=0$ sub-level is represented by $\ket{1}$, while the $m_s=\pm1$ sub-levels, which are degenerate at zero-field, are collectively represented by $\ket{2}$. The triplet ground states $\ket{1}$ and $\ket{2}$ are separated by approximately 2.87~GHz. Optical pumping ($532$~nm) from the $^3A_2$ ground state to a vibronically excited level in the triplet excited manifold $^3E$ rapidly (ps) decays to the electronic excited states, $\ket{3}$ and $\ket{4}$. From here, the spin population can either relax through: (i) radiative relaxation into the triplet ground state with a zero-phonon line (ZPL) of 637~nm and phonon-side band (PSB) that extends to around 800~nm, or (ii) a non-spin-conserving and non-radiative ISC into the singlet upper state $\ket{5}$. The latter transition has a branching ratio of $\sim30\%$ and it is more probable for the $m_s=\pm1$ states than $m_s=0$ state ($k_{45} > k_{35}$). The decay from state $\ket{5}$ to $\ket{6}$ (spin conserving) occurs in less than 1~ns predominantly through non-radiative relaxation, exhibiting a weak ZPL of 1042~nm. \cite{ulbricht2018excited} Similarly, the ISC relaxation from $\ket{6}$ back to $\ket{1}$ occurs at a higher probability than the relaxation to state $\ket{2}$ ($k_{61} > k_{62}$). This leads to spin polarization that enables the NV spin projection to be read out via ODMR. 

In conventional ODMR spectroscopy, the NV state population is determined through changes in the visible PL emission (637 to 800~nm) upon optical and microwave excitation. In singlet infrared absorption ODMR, this is instead achieved by monitoring the transmitted fraction of a 1042~nm laser beam through diamond. This is possible because of the rapid relaxation from $\ket{5}\rightarrow\ket{6}$ ($\tau<1$~ns) that is predominantly non-radiative and the relatively slower ISC, $\ket{6}\rightarrow\ket{1}$ ($\tau\sim100$~ns). \cite{acosta2010optical, tetienne2012magnetic} Optical excitation and microwave driving redistribute the NV population between the triplet ground states, altering the population transferred into the singlet state. This modifies the absorption experienced by a probe laser resonant with the singlet transition. The resulting ODMR signal is therefore observed as a change in the transmitted probe intensity.

When NV-diamond is incorporated into a laser cavity, the singlet absorption acts as an intracavity loss that depends on the spin state populations. Optical and microwave excitation therefore modulates the effective cavity loss, altering the laser operating point and reducing the differential gain of the laser. The relative change in output power exhibits a strong nonlinear relationship as the laser is driven closer towards threshold. Thus, in addition to a geometric enhancement of the absorption path length within the cavity, LTM offers an additional threshold-induced contrast enhancement by operating near the lasing threshold that further amplifies the absorption ODMR contrast. The degree of threshold-induced contrast enhancement depends on the proximity to threshold, cavity gain and loss parameters, and magnitude of the absorption-induced loss modulation. In principle, this provides a route to greatly improved magnetic-field sensitivity.

\section{\label{sec:setup}Experimental setup}

\subsection{Diamond}
The diamond used inside the laser cavity is a $\braket{100}$ high pressure high temperature (HPHT) 1b diamond (Element Six) with dimensions $3.0\times3.0\times0.25$~mm$^3$. The diamond was irradiated with 2~MeV electrons to a total fluence of $8.5\times10^{17}$~cm$^{-2}$ while held at a temperature of 740$\deg$C and a nitrogen ambient at a pressure of $10^{-2}$~mbar for 21~hours. This process produced NV centers uniformly distributed across the entire diamond volume with an estimated NV density of $1.20\pm0.04$~ppm (N density $35.7\pm1.1$~ppm). Detailed processing and NV density characterization can be found in our previous work.\cite{capelli2019increased} 

The inhomogeneous dephasing time $T_2^*$ was estimated from the linewidth of the ODMR resonance feature measured under an applied bias field ($B_0\approx6$~mT) with all eight resonances fully resolved. Lorentzian fits to a single ODMR resonance give $\Delta f=6.6\pm0.5$~MHz and $T_2^*=\frac{1}{\pi \Delta f}=48\pm3$~ns.  

Anti-reflection (AR) and high-reflection (HR) coatings were deposited on the diamond using electron beam evaporation. Prior to coating the diamond, it was first cleaned sequentially with acetone, IPA and DI water in an ultrasonic bath for 5~minutes each and dried with dry nitrogen. The diamond was subsequently loaded into the evaporator chamber and evacuated to a base pressure of $4\times10^{-7}$~mbar. For the AR coating, a $178\pm1$~nm silicon dioxide (SiO$_2$) layer was first evaporated on both diamond surfaces (thickness confirmed by ellipsometry). The SiO$_2$ thickness satisfies the quarter-wavelength condition at 1042~nm, resulting in a measured diamond transmittance of $T>98\%$ at 1042~nm ($R<1\%$ per surface). \cite{yeung2012anti} A titanium/gold bilayer (2/20~nm) was subsequently deposited on one surface to form the HR-coated facet ($R=90\pm2\%$ at 1042~nm). Gold was chosen due to its high reflectivity in the infrared and relatively lower reflectivity in the visible wavelength region. This allows a fraction of the externally introduced 532~nm pump light to pass through and excite the diamond NVs. Additional details of the coating design and measurements of its transmittance are provided in the Supplementary Material (Fig.~S1). \footnote{See Supplemental Material at https://link.aps.org/xxx/ for reflectance and transmittance spectra of thin film coated diamond; effective reflectivity calculations; ODMR parameter sweeps; optical bistability at the laser threshold; P-I curve fitting; modeling ODMR contrast enhancement near the laser threshold; probe laser noise linear spectral density; estimating the number of NV centers in the sensing volume.}

Using the measured NV density (1.2~ppm) and the calculated pump-probe overlap volume, we estimate that approximately $8\times10^9$ NV centers reside within the sensing volume (more details in the Supplementary Material \cite{Note1}). Since the beam intensity decreases substantially away from the beam waist, this value represents an upper bound on the number of NV centers residing within the geometric sensing volume, rather than a direct estimate of the effective number contributing uniformly to the ODMR signal.

\subsection{\label{sec:ecdl}Modified ECDL}

\begin{figure}[t]
\includegraphics[width=1.0\linewidth]{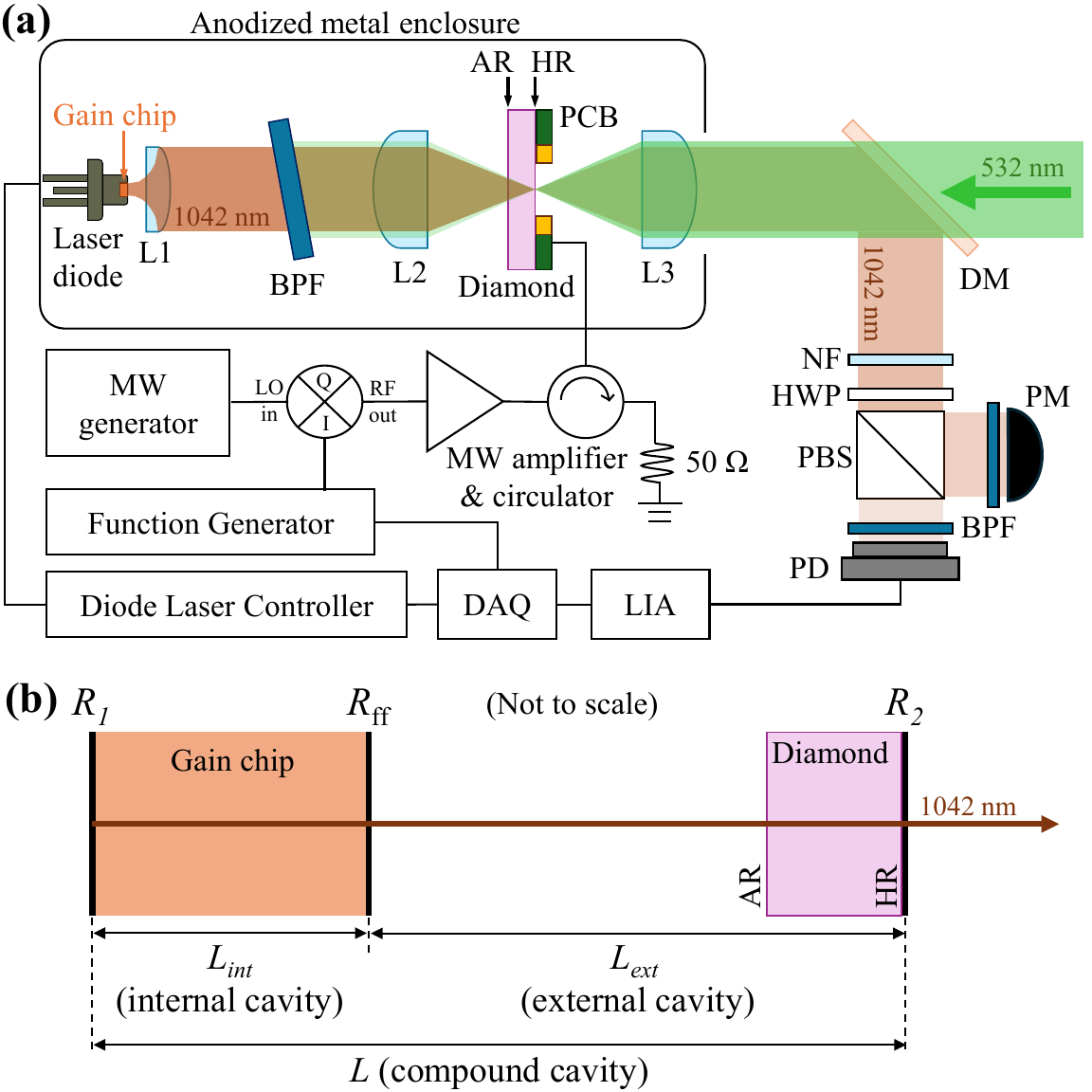}
\caption{\label{fig:setup} (a) Laser cavity setup for diamond NV singlet infrared absorption ODMR. L1: collimating lens. BPF: band pass filter. L2, L3: focusing lens ($f=4.55$ mm, NA$=0.55$). AR: anti-reflection coating. HR: high reflection coating. MW: Microwave. PCB: Printed circuit board (for microwave delivery). DM: Dichroic mirror (short pass 950~nm). NF: notch filter (532~nm). BPF: band pass filter (1050/25~nm). HWP: half wave plate. PBS: Polarizing beam splitter. PM: power meter. PD: photodetector. DAQ: data acquisition device. (b) Schematic of a three mirror model that describes our ECDL. The mirrors are represented by the solid black lines and are indicated as $R_1$, $R_{\rm{ff}}$ and $R_2$. See main text for details. }
\end{figure}

The ECDL used in this work is based on a commercial cat-eye ECDL system (MOGLabs). Here, we briefly describe the key components of the modified configuration, while a more detailed description of the original design can be found in Ref.~\citenum{thompson2012narrow}. A schematic of our ECDL and ODMR experiment is shown in Fig.~\ref{fig:setup}(a).

The gain component is a semiconductor laser diode (Toptica Eagleyard) with an emission range of $980-1090$~nm and an emission center at 1060~nm when operated at 25$\deg$C. The front facet of the gain chip had an AR coating to $R_{\text{ff}}<0.1\%$ and the rear facet highly reflective with $R_1=90\%$ as specified by the manufacturer. Light emitted from the laser diode is collimated by an aspheric lens L1 placed immediately after the diode. The external cavity output mirror (output coupler, OC) is formed by the HR-coated surface of our NV-diamond, which has a reflectivity of $R_2=90\%$ at 1042~nm. The bulk of the diamond and the AR-coated facet are located inside the external cavity. We note that this configuration differs from existing LTM setups where the diamond is placed at Brewster's angle in a laser cavity. The external cavity length is $L_{ext}=30\pm3$~mm. The internal cavity length is $L_{int}=1.5$~mm, as quoted by the manufacturer. The total physical length of the compound cavity is thus $L=L_{int}+L_{ext}=32\pm3$~mm. 

A key feature of our ECDL is the cat-eye OC configuration. As shown in Fig.~\ref{fig:setup}(b), an intracavity focusing lens, L2 (numerical aperture of $\text{NA}=0.55$), focuses the collimated diode emission onto the HR surface of the diamond. The estimated beam waist diameter at the focal point of the lens, expected to coincide with the HR-coated diamond surface, is $2w_0=4\ \mu$m. The ECDL output is re-collimated using an identical lens, L3, positioned after the diamond OC.

At an operating temperature of $T_{\rm{set}}=18\deg$C, the ECDL was tuned to $1042.0$~nm by rotating the filter angle relative to the incident beam while monitoring the wavelength using a MOGLabs Fizeau wavemeter. The FWHM of the laser emission line is estimated at $<100$~kHz. \cite{thompson2012narrow} The laser diode current and temperature are controlled using a MOGLabs diode laser controller (DLC), which includes an ultra-low noise diode current source ($< 100$ pA/$\text{Hz}^{-1/2}$ from DC to 1~MHz), a temperature controller, and a scan generator that provides the option to modulate the diode current source (sawtooth ramp function, $4-37$ Hz). 

Similar to Ref.~\citenum{webb2021laser}, we used a three-mirror model to describe our cavity. This is shown in Fig.~\ref{fig:setup}(b). This compound cavity may be simplified further to two mirrors with reflectivity $R_1$ and $R_{e}$, where the latter represents the effective reflectivity of the diode front facet and external mirror subsystem. We use the expression below for $R_{e}$: 

\begin{align}
    R_{\rm{e}} &= \left| \frac{\sqrt{R_{\rm{ff}}} + \eta_{\rm{overlap}}\sqrt{R_2}}{1 + \eta_{\rm{overlap}}\sqrt{R_{\rm{ff}} R_2}} \right|^2 \label{eq:Reff}
\end{align}

\noindent where $\eta_{\rm{overlap}}$ is the spatial mode overlap efficiency between the internal and external cavity optical fields. Assuming that $\eta_{\rm{overlap}}=0.8$, we obtain $R_{e}=60\%$ (see Supplementary Material \cite{Note1} Section SM-II for further explanation). 

Although the finesse of the active cavity during laser operation is complex to quantify due to the presence of gain, the passive (unpumped) cavity finesse provides a useful baseline for the geometric properties of the system. For the compound cavity without gain, the finesse is given by $F=(\pi(R_1 R_{\rm{e}})^{1/4})/(1-(R_1 R_{\rm{e}})^{1/2})=10$. \cite{yariv2007photonics} This corresponds to an average of $N \approx F/\pi = 3$ round trips, \cite{jensen2014cavity} consistent with the low-finesse regime expected for this configuration.

\subsection{Singlet ODMR spectroscopy and magnetometry}
The NV pump laser (Laser Quantum, gem 532~nm) was introduced externally into the ECDL cavity using a pair of steering mirrors. Coarse spatial overlap of the 532~nm pump and 1042~nm probe beams was achieved by adjusting these mirrors. Fine alignment was achieved by performing minor adjustments to these steering mirrors to maximize the lock-in ODMR amplitude. 

A dichroic mirror (short pass 950~nm) was placed at the ECDL output to separate pump and probe beams. A 532~nm notch filter (NF) was inserted in the probe beam path to suppress residual pump light. The probe beam was then divided into two detection arms using a half-wave-plate (HWP) and polarizing beam splitter (PBS). One arm was directed to a power meter (PM) (Thorlabs S130C) to monitor the probe output power. The second arm was directed to an InGaAs switchable gain photodetector (PD) (Thorlabs PDAPC4) and the signal was recorded using a data acquisition device (DAQ) (National Instruments USB-6212). Both detection arms were equipped with 1050/25~nm band pass filters (BPF) to further isolate the probe wavelength.

Microwave excitation was delivered to the diamond using a custom printed circuit board (PCB) with a broadband loop antenna. The PCB was placed in direct contact with the HR-coated facet of the diamond. Microwaves in the $1-4$~GHz range were generated by a microwave source (Windfreak Technologies SynthNVPro), amplified by a low noise amplifier (Mini Circuits ZRL-3500+) and routed through a microwave circulator before being delivered to the PCB and terminated with a $50\ \Omega$ load resistor to ground.  

For singlet ODMR measurements, the microwave signal was amplitude modulated prior to amplification using a function generator (Keysight Technologies) and a quadrature modulator (Texas Instruments). For magnetometry measurements, we performed lock-in ODMR. In lock-in ODMR, the microwave was frequency modulated using the internal reference of a digital lock-in amplifier (LIA) (Stanford Research Labs SR860). The PD voltage output was demodulated by the LIA and recorded with the DAQ. Laser noise and magnetic-field noise measurements were performed by measuring time traces of 100~s duration (400~kHz sampling rate) of either the PD or LIA voltage output with the DAQ, respectively. In all our measurements, the pump power was fixed at $P_{532}=470$~mW (measured before the OC) and the microwave power was fixed at $24.6$~dBm (measured using a spectrum analyzer), unless specified otherwise. These values were chosen to maximize the threshold shifts and ODMR signal without compromising the linewidth (see Supplementary Material Fig.~S2 for the corresponding parameter dependencies \cite{Note1}).

\section{\label{sec:results}Results and discussion} 

\subsection{Laser gain and threshold characteristics}

\begin{figure*}
\includegraphics[width=0.9\linewidth]{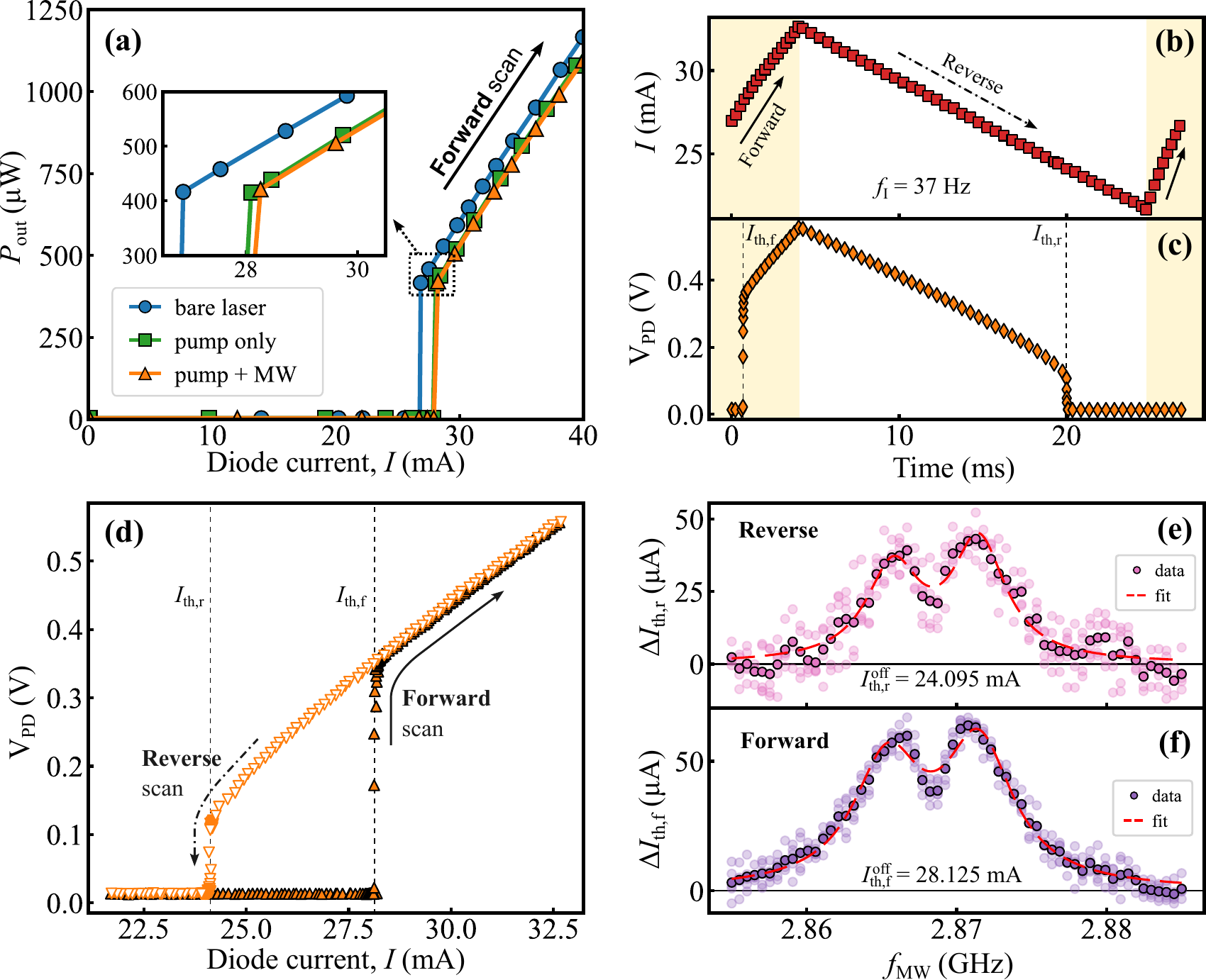}
\caption{\label{fig:pi-curves} (a) Diamond-integrated cat-eye ECDL forward P-I curve measured under various conditions (zero magnetic-field). Solid lines are to guide the eye. Green and orange data points were measured with $P_{532}=300$~mW. (b) Sawtooth waveform that modulates the laser diode current during measurements and (c) the resulting PD voltage. (d) Forward and reverse P-I curves reconstructed from (c) and (d) showing hysteresis in the laser threshold ($I_{\text{th,f}} \neq I_{\text{th,r}}$). NV ODMR spectra obtained by monitoring the (e) reverse and (f) forward laser thresholds. The vertical axes represent the change in $I_{\text{th}}$ relative to the off-resonance value, $I_{\text{th,f/r}}^{\text{off}}$, as indicated. The opaque data points represent the calculated average of the transparent data points (five scans). Red dashed curves are Lorentzian fits to the data.} 
\end{figure*} 

The laser gain characteristics of the modified ECDL were characterized using its power-current (P-I) curve (also referred to as the light-current or L-I curve), shown in Fig.~\ref{fig:pi-curves}(a). The data were acquired by incrementally increasing the diode current via the DLC while simultaneously monitoring the output power, $P_{\text{out}}$, with a calibrated PM (corrected for attenuation by the PBS). The laser exhibits a sharp turn-on at $I=26.75$~mA, where the output power increases by approximately two orders of magnitude from $P_{\rm{out}}=3.49\pm0.01\ \mu$W to $416\pm1\ \mu$W. This occurs when gain from stimulated emission exceeds losses and defines the forward threshold current, $I_{\text{th,f}}$, and threshold power $P_{\rm{th,f}}=416\ \mu$W (blue circles). The observed discontinuity in the laser output arises from optical bistability within the ECDL, the mechanisms of which are detailed in the Supplementary Material. \cite{Note1} Above threshold, $P_{\rm{out}}$ increases linearly with current.

Under simultaneous continuous-wave (cw) 532~nm excitation of the diamond NVs, the P-I curve and laser threshold current were observed to shift towards higher current values, as shown by the green squares in Fig.~\ref{fig:pi-curves}(a). The threshold current increases by 1.33~mA to $I_{\text{th,f}}^{\text{off}}=28.08$~mA while the threshold power remains unchanged. A further, smaller shift is observed under resonant microwave excitation of the NV triplet ground state ($f_{\text{MW}}=2.87$~GHz at zero-field), yielding $I_{\text{th,f}}^{\text{on}}=28.26$~mA (orange triangular data points). These shifts are attributed to an increase in the NV singlet state population and the associated 1042~nm absorption, as discussed below. 

To probe the interaction between the NV and laser cavity, the diode current was modulated using a 37~Hz sawtooth waveform across threshold (Fig.~\ref{fig:pi-curves}(b)), and the corresponding PD response was recorded (Fig.~\ref{fig:pi-curves}(c)). Forward and reverse P-I curves reconstructed from these time traces are shown in Fig.~\ref{fig:pi-curves}(d). A clear hysteresis is observed, with the reverse threshold current ($I_{\mathrm{th,r}} = 24.14$~mA) significantly lower than the forward threshold ($I_{\mathrm{th,f}} = 28.11$~mA). This behavior, observed even without 532~nm pump or microwave excitation, indicates optical bistability of the ECDL. 

The dependence of the threshold current on microwave frequency is shown in Figs.~\ref{fig:pi-curves}(e) and (f) for the reverse and forward scans, respectively. These spectra were obtained by reconstructing P-I curves at each microwave frequency and extracting $I_{\mathrm{th}}$ via geometric fitting with a step function multiplied by a linear function (see Supplementary Material Fig.~S3 \cite{Note1}). The resulting spectra exhibit the characteristic zero-field ODMR lineshape with non-axial crystal strain-induced splitting $2E=4.9\pm0.3$~MHz, \cite{gruber1997scanning, acosta2010temperature} demonstrating that microwave-induced changes in the NV singlet population directly modulate the laser threshold current via absorption losses. Since the forward onset of lasing is sharper and lacks the gradual roll-off observed in the reverse P-I curve, the forward threshold exhibits a higher signal-to-noise ratio (SNR) and can be measured more precisely. 

\subsection{Singlet ODMR}

\begin{figure*}
\includegraphics[width=0.9\linewidth]{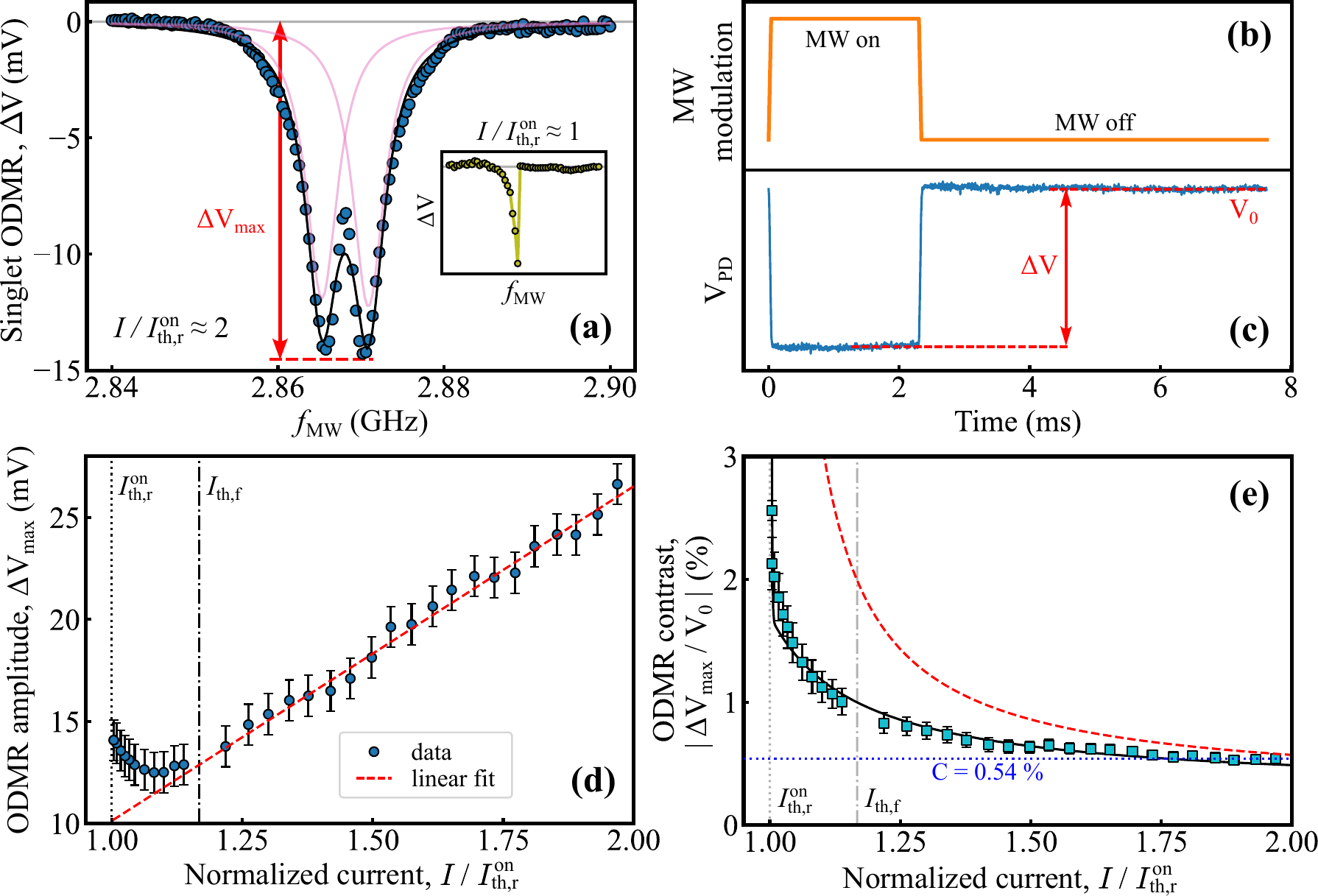}
\caption{\label{fig:odmr} (a) Singlet ODMR spectrum collected at zero magnetic-field and far above the laser threshold. The solid blue circles are the measured data while the solid black and pink curves are Lorentzian fits. Inset: ODMR spectrum collected just above the reverse laser threshold (solid line is to guide the eye). (b) Square wave ($f=131$ Hz, duty cycle $30\%$, single period) that modulates the microwave excitation during ODMR measurements and (c) an example of the measured PD output (10 averages). (d) Maximum ODMR amplitude as a function of the diode current normalized by the reverse, on-resonance threshold current. The forward and reverse threshold current positions are indicated by the vertical lines. The red dashed line is a linear fit to the data in the region $I > I_{\mathrm{th,f}}$. (e) ODMR contrast as a function of the normalized current. The red dashed and solid black curves are the theoretically calculated ODMR contrast based on an ideal and non-ideal model, respectively. The measurements in this figure were performed with $P_{\rm{MW}}=26$~dBm. 
}
\end{figure*}

Figure~\ref{fig:odmr}(a) shows a singlet ODMR spectrum collected far above threshold ($I/I_{\rm{th,r}}^{\rm{on}}\approx2$) by monitoring the change in the probe laser power on the PD while amplitude-modulating the microwave (see Fig.~\ref{fig:odmr}(b) and (c)). The singlet ODMR signal, $\Delta V$, is defined as the difference between the PD voltage with and without microwave applied (red arrow in Fig.~\ref{fig:odmr}(c)). The ODMR spectrum exhibits the strain-induced splitting of $2E=4.6\pm0.1$~MHz with a slightly asymmetric peak amplitude similar to spectra shown in Fig.~\ref{fig:pi-curves}. This asymmetry is most likely due to stray magnetic-fields, \cite{schlussel2018wide} but it may also arise from microwave field inhomogeneities \cite{levchenko2015inhomogeneous} and spin mixing effects. \cite{goldman2015phonon, jin2025first}
The singlet ODMR spectrum collected just above the reverse threshold ($I/I_{\rm{th,r}}^{\rm{on}}\approx1$) is shown in the inset of Fig.~\ref{fig:odmr}(a). At the reverse, on-resonance threshold, the laser switches off completely as the microwave is swept across resonance and the optical bistability prevents the laser from switching back on during the measurement. 

Figure.~\ref{fig:odmr}(d) shows the dependence of the maximum ODMR amplitude, $\Delta V_{\rm{max}}$, on the normalized diode current, $I/I^{\mathrm{on}}_{\mathrm{th,r}}$. We use the reverse, on-resonance threshold value for normalization since it is the minimum threshold value ($I_{\rm{th,r}}^{\rm{on}} < I_{\rm{th,r}}^{\rm{off}} < I_{\rm{th,f}}^{\rm{on}} < I_{\rm{th,f}}^{\rm{off}}$). For $I > I_{\mathrm{th,f}}$, the amplitude increases linearly with current. This is due to an increase in the external cavity absorption loss under on-resonant microwave excitation, thus reducing the slope of the P-I curve as compared to when the microwave is off-resonance. Deviations from linearity occur near threshold and are most pronounced when $I/I^{\mathrm{on}}_{\mathrm{th,r}}<1.05$, where the maximum ODMR amplitude changes from decreasing to increasing towards the laser threshold. This is due to the nonlinear features in the P-I curves near threshold as discussed further below. 

The ODMR contrast, defined as $C = \Delta V_{\mathrm{max}}/V_0$, where $V_0$ is the PD voltage when the microwave is switched off, is shown in Fig.~\ref{fig:odmr}(e). Far above threshold, the contrast saturates at $C_{\rm{far}}=0.54\pm0.03\%$. This represents the cavity enhanced absorption contrast. The single-pass absorption fraction may be estimated through $C_{\rm{single}}=C_{\rm{far}}/(2N)=0.09\%$, where $2N=6$ is the estimated geometric cavity enhancement to the absorption, as calculated in Section~\ref{sec:ecdl}. As the laser approaches the reverse threshold, the contrast increases to $C_{\rm{th}}=2.56\pm0.08\%$, demonstrating a nearly five-fold threshold-induced enhancement to the absorption contrast, or an almost 30-fold enhancement as compared to our single-pass absorption estimate. While this behavior is qualitatively consistent with LTM predictions, \cite{jeske2016laser, dumeige2019infrared, webb2021laser} the degree of threshold-induced contrast enhancement remains far below predictions for ideal LTM systems.

Using a phenomenological model based on Ref.~\citenum{webb2021laser}, we calculate the expected contrast (red dashed line in Fig.~\ref{fig:odmr}(e)) for an ideal laser system (absence of optical bistability and discontinuities in the P-I curves). We then modeled the discontinuity by multiplying the ideal P-I curves with a step function. The calculated contrast of this non-ideal case is shown as a solid black curve. Physically plausible values of the free parameters to the model were selected to obtain agreement with the measured data. While the details of our calculations and fits are shown in the Supplementary Material \cite{Note1} (Fig.~S4), the discrepancy between ideal LTM theory and experiment can be clearly attributed to the non-ideal step feature, which acts to reduce the contrast enhancement near threshold. 

\subsection{Magnetometry and noise}

\begin{figure}
\includegraphics[width=1.0\linewidth]{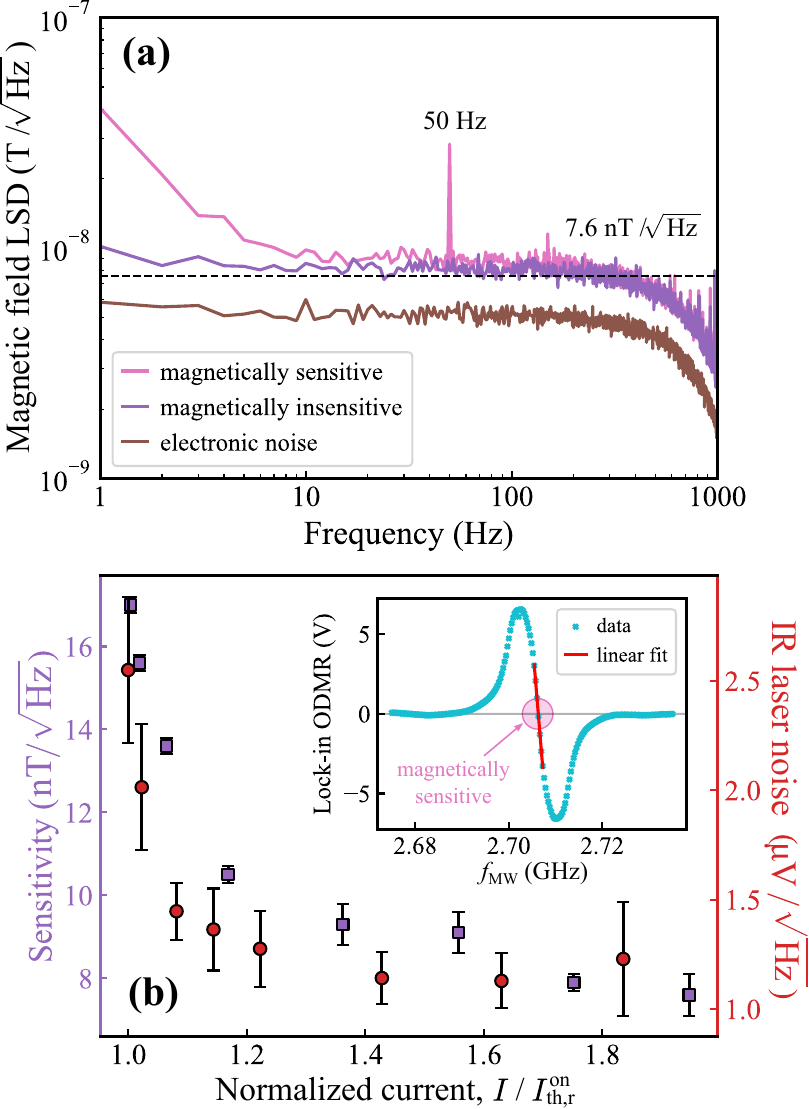}
\caption{\label{fig:mag-noise} (a) Magnetic-field LSD measured far above threshold at $I/I_{\rm{th,r}}^{\rm{on}} \approx 2$. The horizontal dashed line represents the average of the magnetically insensitive spectrum from DC$-500$~Hz. (b) Magnetic-field sensitivity and laser noise as a function of the normalized diode current. Inset: Frequency modulated lock-in ODMR spectrum of a selected resonance measured under an applied bias magnetic-field. }
\end{figure}

Magnetometry measurements were performed under an applied bias magnetic-field to lift the degeneracy of the NV orientations. Frequency-modulated microwave excitation (FM modulation frequency $f_{\rm{mod}}=1.371$~kHz and deviation $\Delta f_{\rm{MW}}=4$~MHz) and lock-in detection were used to measure the ODMR spectrum (first derivative). While all of the eight ODMR resonances were fully resolved, we have selected the resonance at $f_{\rm{MW}}=2.706$~GHz with the largest ODMR amplitude for further measurements (shown in the inset of Fig.~\ref{fig:mag-noise}(b)).

Magnetic-field linear spectral density (LSD) data were obtained by analyzing 100~s time traces of the lock-in output. The data were segmented into 1~s time bins, Fourier transformed, averaged, and converted to magnetic-field units using the slope of the ODMR resonance (linear fit shown in the inset of Fig.~\ref{fig:mag-noise}(b)) and the NV electron gyromagnetic ratio ($\gamma_e = 28.024$~MHz/mT). The resulting noise spectra measured far above threshold, at $I/I_{\rm{th,r}}^{\rm{on}} \approx 2$, are shown in Fig.~\ref{fig:mag-noise}(a). The magnetically sensitive spectrum was collected at the zero-crossing of the lock-in ODMR spectrum (indicated by the pink circle in the inset figure, $f_{\rm{MW}}=2.706$~GHz) while the magnetically insensitive spectrum was measured at a microwave excitation frequency of $f_{\rm{MW}}=1.31$~GHz. The electronic noise spectrum was measured with both the pump and probe lasers blocked from the detector. 

The magnetic-field sensitivity was determined by the average of the magnetically insensitive spectrum from DC$-500$~Hz. This yields a sensitivity of $7.6\pm0.4~\mathrm{nT/\sqrt{Hz}}$ for the spectrum shown in Fig.~\ref{fig:mag-noise}(a). The roll-off in the spectra observed from $\sim1$~kHz is due to the applied LIA filters, which have an equivalent noise bandwidth of 2.6~kHz. In the magnetically sensitive spectrum, low-frequency noise (DC$-10$~Hz) is dominated by stray magnetic-field fluctuations in the laboratory, while a prominent peak at 50~Hz arises from power line interference.

The magnetic-field sensitivity as a function of the normalized diode current is shown in Fig.~\ref{fig:mag-noise}(b). Contrary to idealized LTM predictions, the sensitivity degrades near the laser threshold, reaching $17.0\pm0.2~\mathrm{nT/\sqrt{Hz}}$ at $I/I_{\rm{th,r}}^{\rm{on}}\approx1$ (2.2 times worse than far above the laser threshold). As shown in Fig.~\ref{fig:mag-noise}(b), this degradation arises from the substantial increase in the probe laser noise near the laser threshold, which outweighs the five-fold ODMR contrast enhancement. Relative to operation far above threshold, the threshold-enhanced laser noise degrades the sensitivity at threshold by approximately an order of magnitude. The laser noise data were obtained by analyzing the time traces of the IR laser PD response at different diode currents, and calculating the average noise between $1-5$ kHz in the LSD (Fourier transforms shown in Supplementary Material Fig.~S5 \cite{Note1}). 

The magnetic-field sensitivity of an NV-based magnetometer is fundamentally limited by photon shot noise. In the small-contrast limit, it is given by: \cite{barry2020sensitivity}

\begin{align}
    \eta_B &= \frac{4}{3\sqrt{3}} \frac{h}{g_e \mu_B} \frac{\Delta f}{C\sqrt{R}} \label{eq:sens}
\end{align}

\noindent where $h$ is Planck's constant, $g_e$ is the electron g-factor, $\mu_B$ is the Bohr magneton, $\Delta f$ and $C$ are the ODMR linewidth and contrast, respectively, and $R$ is the detected photon count rate. Using Eq.~\ref{eq:sens} and the results from Fig.~\ref{fig:odmr}(d), we calculate the zero-field, cw-ODMR shot-noise limited sensitivity of our setup to be $520\pm40\ \mathrm{pT/\sqrt{Hz}}$ and $2.4\pm0.2\ \mathrm{nT/\sqrt{Hz}}$ when the laser is operated at and far above threshold, respectively. 

The sensitivities reported in this work are several orders of magnitude poorer than those achieved by state-of-the-art NV magnetometers. It is nevertheless worth mentioning that our measured sensitivities are comparable to those reported in recent works based on the infrared singlet absorption modality of LTM. \cite{wollenberg2025laser, gottesman2024infrared} This suggests that the limitations identified here may be relevant more broadly to this sensing modality.

Two principal factors currently limit the projected shot-noise sensitivity of our system. First, the ODMR contrast is substantially reduced by optical bistability. Second, the HPHT diamond used in this work exhibits a relatively broad ODMR linewidth compared with state-of-the-art CVD-grown diamonds. Using Eq.~\ref{eq:sens}, we estimate that achieving near-unity ODMR contrast by suppressing the optical bistability and reducing the linewidth to 400~kHz through the use of a high-purity CVD diamond would improve the projected near-threshold shot-noise limit from $520~\mathrm{pT/\sqrt{Hz}}$ to $800~\mathrm{fT/\sqrt{Hz}}$. This estimate should be regarded as an optimistic best-case projection, as it assumes an unchanged detected photon rate, ideal ODMR contrast of $C=1$, and negligible additional nonlinear processes near the laser threshold. For sensing protocols and frequency ranges compatible with pulsed control, further improvements may be achievable using pulsed ODMR and dynamical decoupling sequences that extend $T_2^*$. \cite{barry2020sensitivity} Increasing the photon flux is expected to be challenging in LTM systems due to the inherently lower photon fluxes from operating near the laser threshold.

Achieving measured sensitivities approaching the shot-noise limit will require suppressing the threshold-enhanced laser noise and reducing the overall laser noise (technical noise). While some level of reduction in the threshold-enhanced noise may be achievable, it is unlikely to be completely removed. Potential strategies to reduce the overall technical noise include mitigation of threshold nonlinearities and optimization of the laser cavity parameters such as the internal and external cavity mode overlap, cavity finesse, external cavity geometry and thermal stability. We note that common-mode noise rejection is less straightforward in intracavity architectures because the sensing medium forms part of the laser cavity, making it difficult to generate a reference signal that captures laser noise without also containing the magnetic-field signal.

\section{Conclusion}

In conclusion, we have integrated an HPHT NV-diamond into a commercial ECDL and demonstrated singlet infrared absorption ODMR within the laser cavity. We show that the NV-diamond modulates both the laser output power and threshold current and confirm through ODMR measurements that this modulation arises from spin-dependent singlet infrared absorption at 1042~nm. While threshold-current monitoring provides a direct spectroscopic validation of the NV-cavity coupling, establishing its viability as a quantitative magnetometry readout modality requires further investigation. In particular, future work will require active cavity stabilization together with measurements of the noise-equivalent field, magnetic-field sensitivity, and long-term drift characteristics to enable a quantitative comparison with conventional power-readout approaches.

We observed a five-fold enhancement of the singlet ODMR contrast when operating the laser near threshold. However, the accompanying increase in laser noise more than offsets this benefit, resulting in a degraded magnetic-field sensitivity of $17.0\pm0.2\ \rm{nT/\sqrt{Hz}}$ at threshold compared with $7.6\pm0.4\ \rm{nT/\sqrt{Hz}}$ far above threshold (DC$-500$~Hz). Thus, for the present platform, near-threshold operation yields a net negative effect on sensitivity. These results experimentally demonstrate that contrast enhancement alone is insufficient to improve magnetometer performance, and that, in practical LTM implementations, threshold-enhanced contrast must be balanced against threshold-enhanced laser noise.

To approach the sensitivities achieved in state-of-the-art NV magnetometry, future work must address the optical bistability and other nonlinear dynamics observed near the laser threshold. This is expected to improve both the projected shot-noise limit and the measured sensitivity. Further improvements may be achieved through the use of high-purity CVD-grown diamonds with sub-MHz linewidths, together with reductions in threshold-enhanced and technical laser noise through cavity optimizations. While these improvements remain to be demonstrated experimentally, they provide a quantitative framework for understanding how the performance of ECDL-based laser threshold magnetometers may be improved.

Overall, this work establishes the feasibility of using compact, electrically driven ECDL platforms for NV magnetometry, offering a lower power requirement for driving the laser gain medium together with a mechanically robust cavity architecture. By experimentally confirming the contrast enhancement when operating the laser near threshold while simultaneously identifying threshold-enhanced noise as a critical physical limitation, this work provides new insight into the trade-offs governing LTM and identifies the performance bottlenecks that must be addressed in future implementations.

\section*{Acknowledgments}

This research was supported by the Australian Government through the Australian Research Council Discovery Projects funding scheme (DP220103181). The diamond AR and HR coatings were fabricated at the RMIT Micro Nano Research Facility (MNRF) in the Victorian Node of the Australian National Fabrication Facility (ANFF). We acknowledge MOGLabs for in-kind support. S. Q. L. thanks Jess Miller, Marco Capelli, Christopher Lew, Harini Hapuarachchi and Florian Schall for useful discussions. J. J. acknowledges funding from the German Bundesministerium für Bildung und Forschung project NeuroQ, grant no. 13N16485. H. E.-H. acknowledges a South Australian Government Future Industry Making Fellowship.

\section*{Data availability}

The data that support the findings of this article are openly available. \cite{data2026}

\bibliography{refs}

\end{document}